\documentclass[pra,a4paper,superscriptaddress,twocolumn,amsmath,amssymb, longbibliography]{revtex4-1}
\pdfoutput=1
\usepackage[english]{babel}
\usepackage{graphicx}% Include figure files
\graphicspath{{./Figs/}}
\usepackage{color}
\usepackage{bm}% bold math
\usepackage{epstopdf}

\usepackage{verbatim}
\begin{document}
	
\title{Solid-state qubits}
	
\author{M.~Fernando Gonzalez-Zalba}
\email{mg507@cam.ac.uk}
\affiliation{Hitachi Cambridge Laboratory, J. J. Thomson Avenue, Cambridge CB3 0HE, United Kingdom}	
%%%%%%%%%%%%%%%%%%%%%%%%%%%%%%%%%%%%%%%%%%%%%%%%%%%%%%%
	
%\begin{abstract}
%
%\end{abstract}

\maketitle

%%%%%%%%%%%%%%%%%%%%%%%%%%%%%%%%%%%%%%%%%%%%%%%%%%%%%%%
\section{Introduction} 

The quantum world is fascinating. It presents a description of nature that defies our most rooted concepts about what reality is. For example, quantum objects possess \lq\lq spooky\rq\rq\ properties that allow them to be in multiple places at the same time, to move in different directions simultaneously, or to exist and not to exist. In other words, to live several parallel stories. Making use of these parallel stories to compute is the realm of quantum computation, a world in which information is processed using the laws of quantum physics. Nowadays, quantum computers are an object of an extensive experimental and theoretical research, since it is known that, on paper, these can solve many of the current computational challenges. These are the problems that even the most sophisticated supercomputers of today, or tomorrow, will not be able to solve.

For example, quantum computers promise to help us design new materials more efficiently, such as room-temperature superconductors or catalysts for the reduction of greenhouse gases and efficient production of fertilizers. We could do rapid searches in unstructured databases, useful in the era of Big Data and in the identification of genetic diseases. They would also allow us to solve complex optimization problems that would facilitate weather and financial market forecasting. And finally, with them we could decode cryptic codes indecipherable with modern methods~\cite{Montanaro2016}.

Quantum computers use a new type of information unit --- the quantum bit or qubit --- which can be in a quantum superposition of binary states, 0 and 1. To execute some of the simpler versions of the aforementioned quantum algorithms, it is believed processors with about 100 qubits will be necessary and in some cases, to run the complete version will require a processor with millions of qubits. To date, with the most advanced quantum technologies, it has been possible to manufacture universal quantum processors with a maximum of 14 to 17 qubits~\cite{Monz2011,IBM}. That is why the most transcendental challenge facing quantum computing is to increase the number of qubits in quantum processors.

It is here that solid-state qubits promise to play a very remarkable role. Solid-state qubits are a subset of qubits in which the system that codifies the quantum bit is integrated into a solid material. The nanofabrication techniques are very similar to those of the microelectronics industry which allows integrating multiple qubits in millimeter size chips. In this Article, which should not be considered comprehensive, I review the main solid-state quantum computing technologies. I focus mainly on semiconductor and superconductor qubit technologies that present a more advanced level of experimental development. The two share the technological challenge of controlling a quantum system in a quasi-perfect way --- initialization, manipulation, reading the information of each qubit and control of the interaction between neighboring qubits --- and at the same time protecting it from decoherence phenomena --- unwanted interactions with the environment. It is outside the scope of this Article to evaluate topological quantum computation, based on quantum non-Abelian phases of the matter and whose hardware promises to be immune to calculation errors and decoherence. Semiconductor and superconductor technologies are in the early stages of development, but are already producing promising results that I describe below.

\begin{figure*}[htbp]
	\centering
		\includegraphics{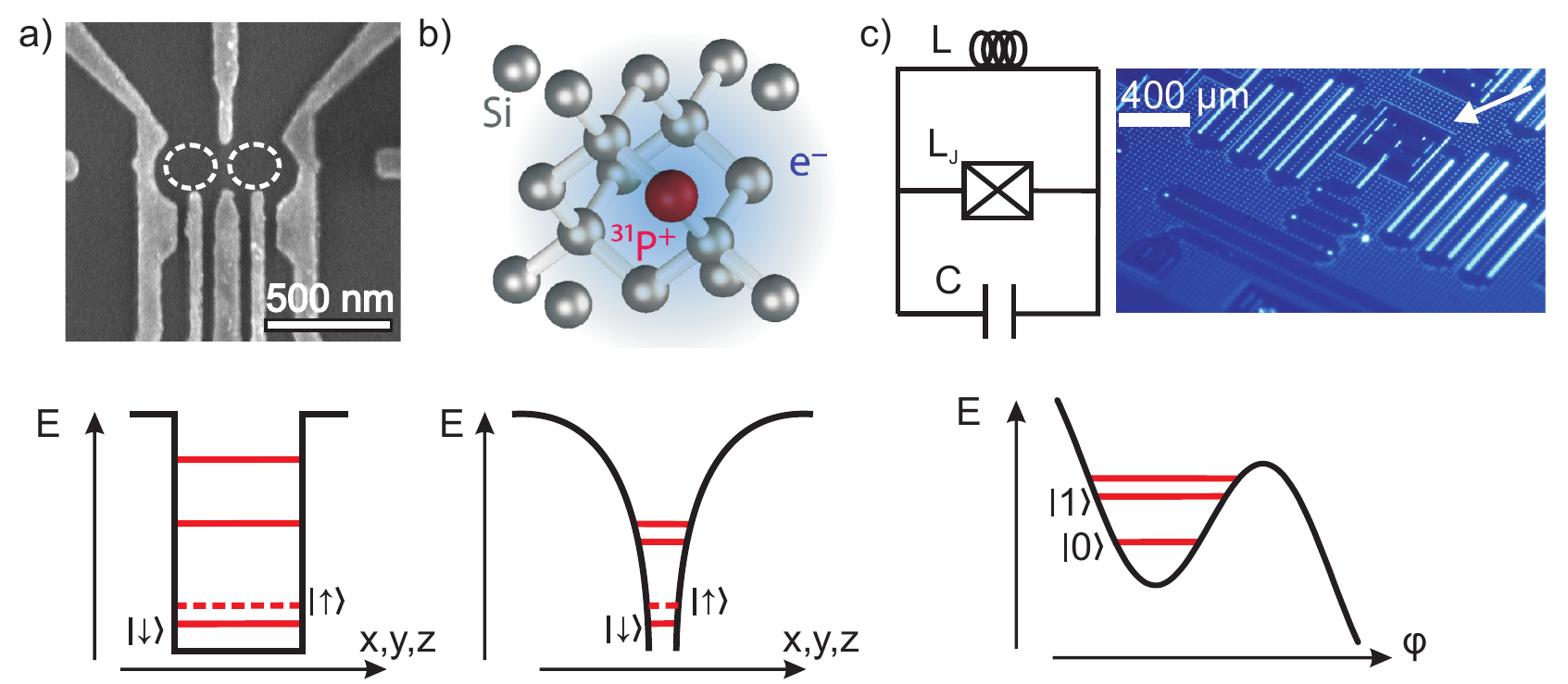}
	\caption{Solid-state qubits and associated potential wells. A) (Top) Scanning electron microscope photograph of a gate-defined double quantum dot in GaAs. The dashed circle indicates the position of the quantum dots. Courtesy of J. Waldie, Semiconductor Physics Group, Cavendish Laboratory. (Bottom) Energy as a function of the spatial coordinates in which the two spin states, $\left|\downarrow\right\rangle$ and $\left|\uparrow\right\rangle$ appear alongside additional excited states. (b) (Top) Schematic of the silicon lattice (lattice parameter $\sim 0.543~$nm) in which a $^{31}$P$^+$ phosphorus impurity and the electron wave function (not to scale) are represented. (Bottom) Potential well of the impurity. (c) (Top) Schematic and optical photograph of a superconducting qubit (indicated by the white arrow). Courtesy of A. D. C\'orcoles, IBM Yorktown Heights. $L$ and $C$ are the inductance and capacitance of the qubit and $L_\text{J}$ corresponds to the inductance of the Josephson junction. (Bottom) Potential well of a phase qubit as a function of the phase difference through the Joshepson junction. $\left|0\right\rangle$ y $\left|1\right\rangle$ are the computational states}.
	\label{fig1}
\end{figure*}

\section{Semiconductor qubits}

Within the set of solid-state qubits that could potentially be integrated on a large scale, we have semiconductor qubits. This type of qubits are based on concepts developed for the microelectronics industry, such as the field effect and semiconductor doping, and use them to process information.

The implementation of the qubit concept in a semiconductor leads naturally to the use of the spin degree of freedom. For example, the spin 1/2 of an electron, in a magnetic field, is a quantum two-level system. To suppress the interaction with phonons, the main source of decoherence in semiconductors, the spins need to be confined and kept at very low temperatures. The confinement source may be artificial, using quantum dots --- nanofabricated electric potential traps (Fig.~\ref{fig1}(a)) --- or natural, using the attractive potential of individual impurities or even its own nuclear spin (Fig.~\ref{fig1}(b)).

In the case of individual impurities, a solid-state qubit that has attracted much attention is the nitrogen-vacancy center in diamond (NV center). The confinement is produced by a nitrogen impurity in close proximity to a vacancy in the carbon lattice. Due to the low mass of the neighboring carbon atoms, decoherence by spin-orbit interaction is weak resulting in a coherence time --- or quantum memory time --- of the electron spin of up to thousands of miliseconds even at room temperature. However, processing diamond in to integrated devices is extremely complex and therefore is currently not considered an ideal candidate for building a large sacle quantum computer. However, NV centers can be used, for example, in high-precision magnetic field sensors or in individual photon sources~\cite{Doherty2013}.

In the case of quantum dots, these can be nanofabricated in two different ways, either by deformation at the interface of epitaxially grown semiconductor heterostructures (self-assembled quantum dots) or induced by the field-effect generated by lithographically-placed gate electrodes on the surface of a semiconductor chip (gate-defined quantum dots). The self-assembled quantum dots allow extremely precise control of single-electron (or hole) spins through optical techniques and their link to individual photons makes them perfect components in future quantum communication technologies. However, randomness in the manufacturing process makes them difficult to integrate on a large scale. On the contrary, gate-defined quantum dots can be fabricated at will and it is possible to design integrated quantum circuits with them. In fact, gate-defined quantum dots are the basis of one of the most famous quantum computation proposals; the Loss and DiVincenzo proposal~\cite{Loss1998}.

The first demonstrations of spin qubits in semiconductors date back to 2005: electronic 1/2 spins in gate-defined quantum dots in GaAs/AlGaAs heterostructures. In quantum dots, the quantum information (in this case the spin state) can be initialized using spin-dependent tunneling, manipulated using electron spin resonance techniques or exchange interaction and read employing high precision electrometers such as quantum point contacts or single-electron transistors~\cite{Hanson2007}. However, the intrinsic coherence time of electronic spins in GaAs turns out to be too short for technological applications in quantum computation. The GaAs nuclear spin bath interacts in an uncontrolled fashion with the target spin qubit through the hyperfine interaction, rapidly destroying its coherence on a time-scale of only 10 ns. Taking into account that in GaAs the time needed to rotate a spin with the exchange interaction is about 1 ns, only about 10 operations can be performed before the quantum information is lost. The most recent research on GaAs qubits has focused on developing dynamic decoupling techniques that have extended the coherence time long enough to demonstrate the interaction between two qubits. Although in GaAs, it is extremely complex to build multi-qubit processors, due to their very short coherence times, this material has been a very useful test bench for experimenting with different spin qubits schemes and manipulation protocols.

Since 2011, research on semiconductor qubits focuses on low nuclear spin materials, such silicon or Si/SiGe heterostructures. In silicon-based devices, the situation is different than in GaAs, since the most abundant isotope, $^{28}$Si, has zero nuclear spin, which avoids undesired hyperfine couplings. Natural silicon contains only 4,7\% of $^{29}$Si, the only stable isotope with nuclear spin other than zero and this percentage can be further reduced by isotope purification techniques.

Early efforts to demonstrate spin qubits in silicon focused on the natural confinement offered by individual impurities (Fig.~\ref{fig1}(b)). In particular, using the electron and nuclear spin of a single phosphorus atom in silicon, as Bruce Kane suggested in his famous proposal~\cite{Kane1998}. In 2014, using isotopically purified silicon, researchers from the University of New South Wales (UNSW) demonstrated electron and nuclear spin qubits with coherence times of 270~$\mu$s and 600~ms respectively [8], the latter being the longest coherent time measured in a solid state device to date. This important improvement allows to make more than 500 operations within the coherence time, all done with an excellent precision in the initialization, manipulation and reading processes. However, this technology suffers a major drawback: impurities need to be positioned in the silicon lattice with atomic precision to produce predictable interactions between qubits. At present, the single ion implantation techniques do not allow to reach this of level precision --- they are limited to 10 nm --- which leads to the difficulty to demonstrate interactions between two qubits. Multi-qubit architectures can only be manufactured with the precision of the scanning probe lithography, a technique that still requires further development~\cite{Fuechsle2012}.

The last major achievement among semiconductor qubits occurred in 2014, when UNSW researchers, combining the benefits of isotopically purified silicon with the versatility in the manufacturing of gate-defined quantum dots, demonstrated spin qubits with manipulation and coherence times of 1~$\mu$s and 120~$\mu$s respectively. The latter can be extended up to 28~ms using decoupling techniques, making it one of the most coherent systems in nature. Although, the manipulation time by electron spin resonance is relatively slow, the fidelity in the control is very similar to that of impurity-based qubits. Finally, this same group in 2015, demonstrated the most complex semiconductor-based quantum circuit to date, in which two qubits interact in a controlled way using a Controlled-PHASE gate~\cite{Veldhorst2015}.

\section{Superconductor qubits}

On the other hand, we have superconducting qubits, which present a more advanced level of technological development than semiconductor qubits. They represent one of the most interesting and promising options to build a quantum computer. These systems, operated at very low temperatures, offer the ability to manipulate their energy spectrum in an extremely precise way and, in fact, sometimes, like quantum dots, they are called \lq\lq artificial atoms\rq\rq\ because of the inherent versatility for manufacturing. Unlike atoms and quantum dots, superconducting qubits are almost macroscopic structures, with physical dimensions ranging up to the hundreds of microns, which facilitates manipulation and coupling between qubits (Fig.~\ref{fig1}(c)). At the same time, and due to their large dimension and stronger coupling to their environment, the coherence times are lower than those observed in other systems as quantum dots and impurities.

Superconducting qubits are essentially non-linear LC oscillators. Manufactured on a sapphire or silicon substrate, the superconducting materials used in the most advanced samples are mainly niobium and aluminum, with some notable examples in ceramic compounds such as titanium nitride. The nonlinear element that is used in these qubits is a component that, while non-linear, is non-dissipative: the Josephson junction. Josephson junctions consist of two superconductors (usually aluminum) separated by a small insulation layer (Al$_2$O$_3$) of a thickness of the order of 1~nm. These structures are what make possible the selective excitation of a single transition in the entire energy spectrum of the system, something impossible in the harmonic spectrum of an LC circuit.

The first experimental demonstration of a superconductor qubit was attributed to NEC Corporation in 1999~\cite{Nakamura99}. Although with an extremely limited coherence time (1~ns), this experiment promoted to the development of superconductor-based devices in laboratories around the world.

\begin{figure}
	\centering
		\includegraphics{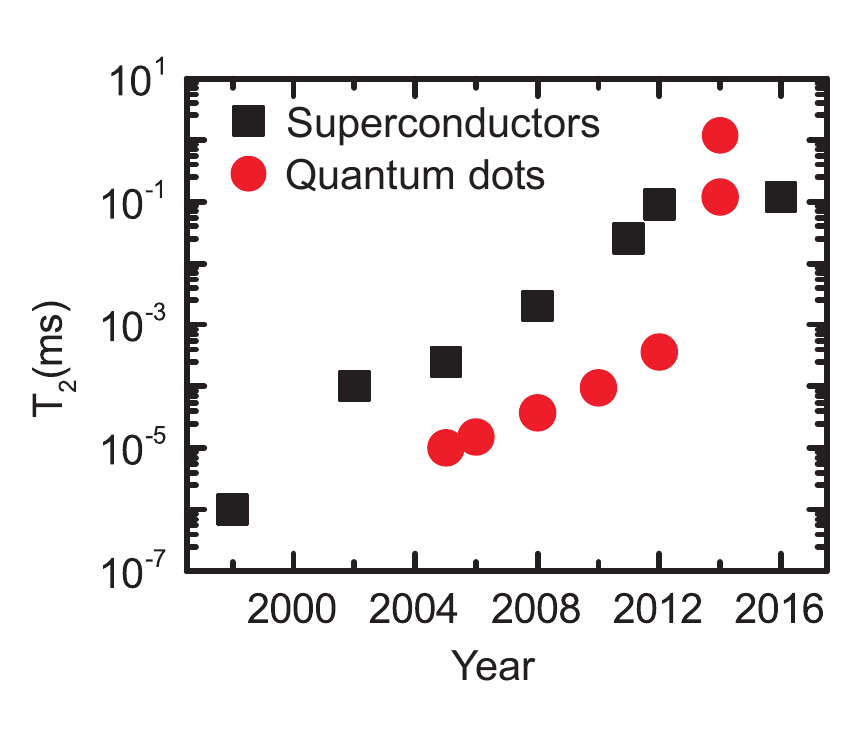}
	\caption{Coherence times of superconducting qubits (black squares) and quantum-dot based qubits (red circles) as function of year of demonstration.}
	\label{fig2}
\end{figure}

During the first years of experimental development of these systems, a classification consisting of three large groups of superconducting qubits was established depending on the way the quantum information was coded. Using the number of Cooper pairs (charge qubit), the direction of a supercurrent around a loop (flux qubit) or the oscillatory state of the circuit (phase qubit). Many designs today are based on variations of these three types, giving rise to more optimized structures such as the Xmon, the transmon or the fluxonium~\cite{Devoret2013}. These qubits can be controlled with microwave signals, voltages, magnetic fields and electric currents in times of about 10~ns. Initially, superconducting qubits were extremely sensitive to decoherence mechanisms, such as charge noise, magnetic flux noise, or dissipation of energy in the dielectric material of the substrate. With time, numerous advances in the study of materials, fabrication, designs and dynamic decoupling techniques have improved this technology by increasing the coherence time by more than five orders of magnitude (Fig.~\ref{fig2}) until reaching 120~$\mu$s in the IBM Quantum Experience in 2016~\cite{IBM}.

To manipulate and read the state of the qubits, while facilitating the coupling between distant qubits, superconductors use a circuit-adapted version of the successful concept of cavity quantum electrodynamics, initially conceived for the study of the interaction between radiation and matter. In circuit quantum electrodynamics, the optical cavity is replaced by a superconducting electrical resonator and atoms by qubits. In the dispersive limit, when the coupling energy between the oscillator and the qubit is much smaller than the difference between ground state energies, the qubit quantum state can be inferred from the resonator response to an electromagnetic pulse without destroying the qubit quantum state. This type of resonators, which can be coplanar wave guides or macroscopic cavities made of aluminum, are extremely versatile, since, in addition to being used to extract information from the qubits, it is often used for individual qubit manipulation and for multi-qubit coupling (see Fig.~\ref{fig1}(c)).
 
Parallel to the development and improvement of coherence in superconducting systems, a variety of two-qubit interactions with different approaches have been developed. For fixed-frequency qubits, cross-resonance~\cite{Rigetti10} or an oscillator-induced phase gate~\cite{Cross15} can be used. For variable-frequency qubits, the controlled-phase gate is often used which uses the higher energy levels of the qubit to exploit entanglement~\cite{Barends14}.

The advance of this technology in the last decade is so spectacular that today superconductor-based quantum processors of up to 9 and 17 qubits~\cite{Barends14, IBM}. It has also promoted the search for quantitative error correction protocols that maintain the processor coherent indefinitely. Noteworthy are, the surface code~\cite{Fowler2012}, which already has several very interesting demonstrations~\cite{Corcoles15, Riste15}, or the so-called \lq\lq cat codes\rq\rq, which use non-linear oscillators to encode the equivalent quantum information of several qubits in a single resonant cavity~\cite{Leghtas13}. 

\section{Conclusions and Future}

Today it is not known what will be the technology with which a fully-fledged quantum computer will be built but what it is clear is that solid-state qubits are one of the most promising candidates. In fact, researchers believe that the quantum computer of the future will probably be a hybrid of different technologies. Just as in a common computer the CPU is composed mostly of transistors and the hard disk by multilayer magnetic materials, quantum dots in silicon with their long coherence times could be the basis of the quantum memory, while superconductors, with their processing speed, could be the quantum information processing unit (QPU). This hybridization between technologies could occur not only at the architecture level, but also at the qubit level. Recent studies have shown that a new type of qubits, known as topological qubits, may exist, for example, in semiconductors with strong spin-orbit interaction when located near superconductors~\cite{Mourik1003}. These qubits are constructed through exotic excitations, called Majorana particles that present non-commutative quantum statistics and allow the storage and manipulation of quantum information in a non-local way. This non-local character protects the topological qubits from imperfections in control protocols and makes them extremely robust against decoherence phenomena.

With the prospect of a universal system with complete error-correction relatively distant in time, nowadays a question has become very relevant: When could we build a quantum chip that can reach \lq\lq quantum supremacy\rq\rq ? That is to say, a chip that allows to perform calculations impossible for a conventional computer. We know that the best supercomputers based on transistor logic that we will ever be able to manufacture will have problems to simulate systems of around 50 qubits. This has triggered an industrial race to be the first to build a chip with the necessary features to achieve this level of supremacy. Google and IBM lead the way with their work in superconducting qubits but Intel and Hitachi are making great efforts to develop quantum processors and quantum memories based on silicon metal-oxide-semiconductor technology~\cite{Gonzalez-Zalba2015}. Equally, there is a great interest in developing quantum algorithms that can be run with about 50 qubits and limited error-correction. With processors of a reasonable number of qubits, quantum simulation or optimization problems could soon be solved, using, for example, heuristic methods in a hardware essentially governed by the laws of quantum mechanics~\cite{Barends16}.

%%%%%%%%%%%%%%%%%%%%%%%%%%%%%%%%%%%%%%%%%%%%%%%%%%%%%%%

%\emph{Acknowledgements- }
%We thank so and so for discussions. The samples presented in this work were designed and fabricated by the TOLOP project partners, http://www.tolop.eu. This research is supported by.

%%%%%%%%%%%%%%%%%%%%%%%%%%%%%%%%%%%%%%%%%%%%%%%%%%%%%%%
\bibliographystyle{unsrt}

\begin{thebibliography}{10}

\bibitem{Montanaro2016}
Ashley Montanaro.
\newblock Quantum algorithms: an overview.
\newblock {\em Npj Quantum Information}, 2:15023 2016

\bibitem{Monz2011}
Thomas Monz, Philipp Schindler, Julio~T. Barreiro, Michael Chwalla, Daniel
  Nigg, William~A. Coish, Maximilian Harlander, Wolfgang H\"ansel, Markus
  Hennrich, and Rainer Blatt.
\newblock 14-qubit entanglement: Creation and coherence.
\newblock {\em Phys. Rev. Lett.}, 106:130506, 2011.

\bibitem{IBM}
The~IBM quantum~experience. http://www.research.ibm.com/ibm{-}q.

\bibitem{Doherty2013}
Marcus~W. Doherty, Neil~B. Manson, Paul Delaney, Fedor Jelezko, Jörg
  Wrachtrup, and Lloyd~C.L. Hollenberg.
\newblock The nitrogen-vacancy colour centre in diamond.
\newblock {\em Physics Reports}, 528(1):1, 2013.

\bibitem{Loss1998}
Daniel Loss and David~P. DiVincenzo.
\newblock Quantum computation with quantum dots.
\newblock {\em Phys. Rev. A}, 57:120--126, 1998.

\bibitem{Hanson2007}
R.~Hanson, L.~P. Kouwenhoven, J.~R. Petta, S.~Tarucha, and L.~M.~K.
  Vandersypen.
\newblock Spins in few-electron quantum dots.
\newblock {\em Rev. Mod. Phys.},79:1217--1265, 2007.

\bibitem{Kane1998}
B.~E. Kane.
\newblock A silicon-based nuclear spin quantum computer.
\newblock {\em Nature}, 393:133--137, 1998.

\bibitem{Fuechsle2012}
Martin Fuechsle, Jill~A. Miwa, Suddhasatta Mahapatra, Hoon Ryu, Sunhee Lee,
  Oliver Warschkow, Lloyd C.~L. Hollenberg, Gerhard Klimeck, and Michelle~Y.
  Simmons.
\newblock A single-atom transistor.
\newblock {\em Nat. Nanotech.}, 7:242--246, 2012.

\bibitem{Veldhorst2015}
M.~Veldhorst, C.~H. Yang, J.~C.~C. Hwang, W.~Huang, J.~P. Dehollain, J.~T.
  Muhonen, S.~Simmons, A.~Laucht, F.~E. Hudson, K.~M. Itoh, A.~Morello, and
  A.~S. Dzurak.
\newblock A two-qubit logic gate in silicon.
\newblock {\em Nature}, 526:410--414, 2015.

\bibitem{Nakamura99}
Y.~Nakamura, Yu.~A. Pashkin, and J.~S. Tsai.
\newblock Coherent control of macroscopic quantum states in a
  single-cooper-pair box.
\newblock {\em Nature}, 398(6730):786--788, 1999.

\bibitem{Devoret2013}
M.~H. Devoret and R.~J. Schoelkopf.
\newblock Superconducting circuits for quantum information: An outlook.
\newblock {\em Science}, 339(6124):1169--1174, 2013.

\bibitem{Rigetti10}
Chad Rigetti and Michel Devoret.
\newblock Fully microwave-tunable universal gates in superconducting qubits
  with linear couplings and fixed transition frequencies.
\newblock {\em Phys. Rev. B}, 81:134507, 2010.

\bibitem{Cross15}
Andrew~W. Cross and Jay~M. Gambetta.
\newblock Optimized pulse shapes for a resonator-induced phase gate.
\newblock {\em Phys. Rev. A}, 91:032325, 2015.

\bibitem{Barends14}
R.~Barends, J.~Kelly, A.~Megrant, A.~Veitia, D.~Sank, E.~Jeffrey, T.~C. White,
  J.~Mutus, A.~G. Fowler, B.~Campbell, Y.~Chen, Z.~Chen, B.~Chiaro,
  A.~Dunsworth, C.~Neill, P.~O'Malley, P.~Roushan, A.~Vainsencher, J.~Wenner,
  A.~N. Korotkov, A.~N. Cleland, and John~M. Martinis.
\newblock Superconducting quantum circuits at the surface code threshold for
  fault tolerance.
\newblock {\em Nature}, 508(7497):500--503, 2014.

\bibitem{Fowler2012}
Austin~G. Fowler, Matteo Mariantoni, John~M. Martinis, and Andrew~N. Cleland.
\newblock Surface codes: Towards practical large-scale quantum computation.
\newblock {\em Phys. Rev. A}, 86:032324, 2012.

\bibitem{Corcoles15}
A.~D. C{\'o}rcoles, Easwar Magesan, Srikanth~J. Srinivasan, Andrew~W. Cross,
  M.~Steffen, Jay~M. Gambetta, and Jerry~M. Chow.
\newblock Demonstration of a quantum error detection code using a square
  lattice of four superconducting qubits.
\newblock {\em Nat. Commun.}, 6:6979 EP --, 04 2015.

\bibitem{Riste15}
D.~Rist{\`e}, S.~Poletto, M.~Z. Huang, A.~Bruno, V.~Vesterinen, O.~P. Saira,
  and L.~DiCarlo.
\newblock Detecting bit-flip errors in a logical qubit using stabilizer
  measurements.
\newblock {\em Nat. Commun.}, 6:6983, 2015.

\bibitem{Leghtas13}
Zaki Leghtas, Gerhard Kirchmair, Brian Vlastakis, Robert~J. Schoelkopf,
  Michel~H. Devoret, and Mazyar Mirrahimi.
\newblock Hardware-efficient autonomous quantum memory protection.
\newblock {\em Phys. Rev. Lett.}, 111:120501, 2013.

\bibitem{Mourik1003}
V.~Mourik, K.~Zuo, S.~M. Frolov, S.~R. Plissard, E.~P. A.~M. Bakkers, and L.~P.
  Kouwenhoven.
\newblock Signatures of Majorana fermions in hybrid
  superconductor-semiconductor nanowire devices.
\newblock {\em Science}, 336(6084):1003--1007, 2012.

\bibitem{Gonzalez-Zalba2015}
M.~F. Gonzalez-Zalba, S.~Barraud, A.~J. Ferguson, and A.~C. Betz.
\newblock Probing the limits of gate-based charge sensing.
\newblock {\em Nat. Commun.}, 6:6084, 2015.

\bibitem{Barends16}
R.~Barends, A.~Shabani, L.~Lamata, J.~Kelly, A.~Mezzacapo, U.~Las Heras,
  R.~Babbush, A.~G. Fowler, B.~Campbell, Yu~Chen, Z.~Chen, B.~Chiaro,
  A.~Dunsworth, E.~Jeffrey, E.~Lucero, A.~Megrant, J.~Y. Mutus, M.~Neeley,
  C.~Neill, P.~J.~J. O'Malley, C.~Quintana, P.~Roushan, D.~Sank,
  A.~Vainsencher, J.~Wenner, T.~C. White, E.~Solano, H.~Neven, and John~M.
  Martinis.
\newblock Digitized adiabatic quantum computing with a superconducting circuit.
\newblock {\em Nature}, 534(7606):222--226, 2016.

\end{thebibliography}

\end{document}